\documentclass[namedreferences]{solarphysics}

\usepackage[hyperref,optionalrh]{spr-sola-addons} 
\usepackage{graphicx}        
\usepackage{amssymb}        
\usepackage{color}           
\usepackage{breakurl}        

\chardef\us=`\_

\begin{document}

\begin{article}
\begin{opening}

\title{Tidally synchronized solar dynamo: a rebuttal}

\author[addressref={aff1},email={Henri-Claude.Nataf@univ-grenoble-alpes.fr}]{\inits{H.-C.}\fnm{Henri-Claude}~\lnm{Nataf}\orcid{0000-0001-6737-1314}}
\address[id=aff1]{Univ Grenobles Alpes, CNRS, ISTerre, Grenoble, France.}

\runningauthor{Nataf H.-C.}
\runningtitle{Tidally synchronized solar dynamo: a rebuttal}


\begin{abstract}
The idea of a planetary origin for the solar cycle dates back to the nineteenth century.
Despite unsurmounted problems, it is still advocated by some.
\citet{stefani2019} thus recently proposed a scenario based on this idea.
A key argument they put forward is evidence that the $\sim11$ years-solar cycle is \textit{clocked}.
Their demonstration rests upon the computation of a ratio proposed by \citet{dicke1978} applied to the solar cycle time series of \citet{schove1955}.
I show that their demonstration is invalid, because the assumptions used by Schove to build his time series force a \textit{clocked} behaviour.
I also show that instabilities in a magnetized fluid can produce fluctuation time series that are close to being \textit{clocked}.
\end{abstract}
\keywords{Solar Cycle, Models; Solar Cycle, Observations}
\end{opening}

\section{Introduction}
     \label{S-Introduction}
     
The recent opening and publication by Courtillot, Le Mou\"el and Lopes (\citealp{malburet2019}) of a \textit{pli cachet\'e} (sealed letter), entrusted to the French Academy of Science by Jean Malburet in 1918 \citep{malburet1918}, highlights the early interest in the search for a link between solar cycles and tides.
In fact, the work of Malburet was already known from a detailed and luminous report he wrote (in french) for the journal \textit{`L'Astronomie'} in 1925 \citep{malburet1925}\footnote{This article is available on line from the Biblioth\`eque de France via Gallica: https://gallica.bnf.fr/ark:/12148/bpt6k9628963x/f369.item.}.
In that report, he correctly estimates the tidal forcing of planets on the Sun as being proportional to $m_p/d_p^3$ where $m_p$ and $d_p$ are the mass of a planet, and its distance from the Sun, respectively.
This scaling yields tidal forcings proportional to $\sim4.0$, $\sim3.8$, $\sim1.8$, and $\sim1.7$, for Jupiter, Venus,  Earth and Mercury, respectively (the contributions of the other planets are at least 10 times smaller).
With this in mind, Malburet shows a good correspondance between the dates of solar maxima and the dates of `weak deviations from Jupiter-Venus-Earth syzygies\footnote{alignment of all these planets with the Sun.}'.
Malburet's idea was later taken up and extended by \citet{wood1972} and \citet{okhlopkov2016}.

There are however two serious problems with this idea:
\begin{enumerate}
\item The amplitude of the tidal forcing on the Sun is extremely small ($<10^{-8}$ kg m$^{-3}$), yielding accelerations one thousand times smaller than observed in the convective zone of the Sun \citep{deJager2005}. The forcing is {\bf 100\,000 times smaller} that the tidal forcing of the four Gallilean satellites on Jupiter (which is similar to the tidal forcing of the Moon on Earth).
\item The $\sim11.2$ years-period inferred from the `weak deviations from Jupiter-Venus-Earth  syzygies' is an artificial construction which has no signature in the complete tidal signal, as demonstrated by \citet{okal1975}, and illustrated in Appendix \ref{A-tide}.
\end{enumerate}

It is therefore surprising that this idea receives a renewed attention \citep{scafetta2012,okhlopkov2016,baidolda2017,courtillot2021,charbonneau2022}. Even more surprising is the enthusiasm shown by Stefani and colleagues who have published no less than 7 articles exploiting this idea (see \citet{stefani2021} and references therein).
It seems that the main reasons that give confidence to these authors is their demonstration that the solar cycle is \textit{clocked}, and probably the belief that this property requires a clocked forcing that only planetary motions can provide.
Their demonstration, exposed in \citet{stefani2019}, rests upon the computation of `Dicke's ratio' of a 1000-years long time series of solar minima, which favors a clocked origin over a random-walk type origin.

The main objective of the present article is to show that the demonstration of \citet{stefani2019} is invalid.
I also show examples of fluid instabilities that naturally produce clocked-looking time series. 

\section{The demonstration of Stefani \textit{et al} (2019)}
\label{S-Stefani}

\citet{stefani2019} picked up an idea that \citet{dicke1978} proposed for testing whether there is a \textit{``chronometer hidden deep in the Sun''}.
The aim of Dicke was to distinguish a clocked behaviour from an \textit{`eruption hypothesis'}, in which solar cycles would appear with a random phase.
While Dicke restricted his analysis to the post-1705 time series of 25 solar maxima, \citet{stefani2019} extend it to 92 solar cycles starting in AD 1000, in an attempt to obtain a better statistical significance.

\subsection{Dicke's ratio}
\label{S-Dicke}

\citet{dicke1978} noticed that a succession of 3 very short cycles starting in 1755 was followed by a very long cycle, as if the Sun was trying to keep up with some internal clock.
He proposed several statistical tools to assess the existence of such a clock.

Consider a time series $t(i)$ of $N$ events $i$.
In a perfectly clocked time series, all event dates $t(i)$ are evenly spaced.
When gaussian distributed noise is added, each event date is displaced from the regular grid by some random time, yielding a corresponding distribution of cycle durations $d(i) = t(i) - t(i-1)$.
In contrast, when events occur with a random phase, cycle durations $d(i)$ have a gaussian distribution, and event dates are obtained as $t(i) = t(i-1)+d(i)$.
Clocked and random-walk time series can yield the same mean cycle duration $\bar{d}$ and standard deviation $\sigma$, but their statistical properties are not all identical.

\citet{dicke1978} introduced a ratio that measures this difference, which \citet{stefani2019} refer to as `Dicke's ratio'.
Dicke's ratio $Di(n)$ computed for subsets of $n \le N$ consecutive events is defined by:

\begin{equation}
Di(n) = \frac{\sum_{i=2}^n \delta_n^2(i)}{\sum_{i=2}^n(\delta_n(i)-\delta_n(i-1))^2},
\end{equation}
where $\delta_n(i)=t(i)-f_n(i)$ is the deviation of event date $t(i)$ from a best linear fit $f_n(i) = a_n i+b_n$ of the $n$ dates.

According to \citet{dicke1978}, the expectation of Dicke's ratio is:

\begin{equation}
\mathbb{E}(Di_{clock}(n)) = \frac{n^2-1}{2(n^2+2n+3)} {\longrightarrow} \frac{1}{2} \textrm{ when } n \to \infty
\label{E-clock}
\end{equation}
for a clocked time series, while it is:

\begin{equation}
\mathbb{E}(Di_{rand}(n)) = \frac{(n+2)(n^2-1)}{3(5n^2+6n-3)} {\longrightarrow} \frac{1}{15}n  \textrm{ when } n \to \infty
\label{E-random}
\end{equation}
for a random-walk time series.

The expectation of $Di(n)$ is independent of $\bar{d}$ and $\sigma$ for both families, but the spread around the expectation does depend upon $\sigma$.
\citet{dicke1978} applied this statistical tool to the time series of sunspot numbers starting in 1705.
Due to the limited number of cycles, he could not reach a very definitive conclusion.

\subsection{Schove's solar cycle time series} 
      \label{S-Schove}      

\begin{figure}    
   \centerline{\includegraphics[width=0.8\textwidth,clip=]{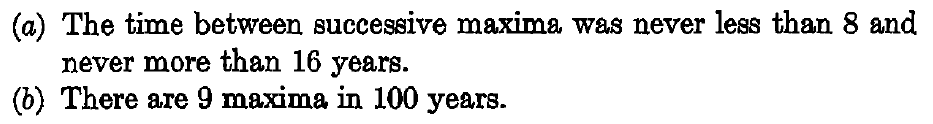}
              }
              \caption{The two assumptions made by \citet{schove1955} to construct his series of solar maxima, as displayed in his article p.131.
                      }
   \label{F-Schove}
   \end{figure}

To reach a firmer conclusion, \citet{stefani2019} complemented the post-1705 solar minima series by solar minima dates as far back as AD 1000, following \citet{schove1955}.
Indeed, Schove published  in 1955 the outcome of a very ambitious venture: dating maxima and minima of the solar cycle from 653 BC to AD 2025!
Pre-1705 observations of sunspots being very rare, he mostly relied on reports of the observation of polar aurorae.
In order to make up for the limited amount of reliable data, \citet{schove1955} explicitly mentions (p.131) that he made use of two {\bf assumptions} to build his 26-century-long table.
These assumptions are reproduced in Figure \ref{F-Schove}.

\section{Arguments for a rebuttal} 
  \label{S-invalidation}

Schove's assumption $(b)$ as listed in Figure \ref{F-Schove} clearly suggests that his time series of solar maxima is {\bf clocked by construction}.
In order to be more specific, I have built synthetic solar cycles to test the impact of Schove's assumptions on the character of the resulting time series.
\subsection{Synthetic solar cycles} 
  \label{S-Synthetic}
  
The well-documented 24 solar cycles from 1755 yield a cycle duration (time between maxima) of $11.0 \pm 2.0$ years.
Extending backwards to AD 1000 with Schove's dates yields 92 maxima separated by $11.1 \pm 2.2$ years.
I have built two different families of synthetic cycles: a random-walk family, and a clocked family.
Both retain the post-1755 dates of solar maxima, as distributed by WDC-SILSO, Royal Observatory of Belgium, Brussels \citep{sidc2022}.

\begin{itemize}
\item The random-walk series are built by drawing at random normally distributed cycle durations, with a mean of $11.1$ years and a standard deviation of $2.0$ years.
The dates of maxima are then constructed by cumulative difference from the date of the oldest post-1755 maximum.

\item The clocked series are built by extending the post-1755 dates of maxima backwards in time with a constant duration of $11.1$ years, and then adding to the obtained dates a normal random noise with zero-mean and a standard deviation of $1.5$ years, this value providing the desired standard deviation of $2.0$ years for cycle durations.
\end{itemize}

  \begin{figure}    
   \centerline{\hspace*{0.015\textwidth}
               \includegraphics[width=0.415\textwidth,clip=]{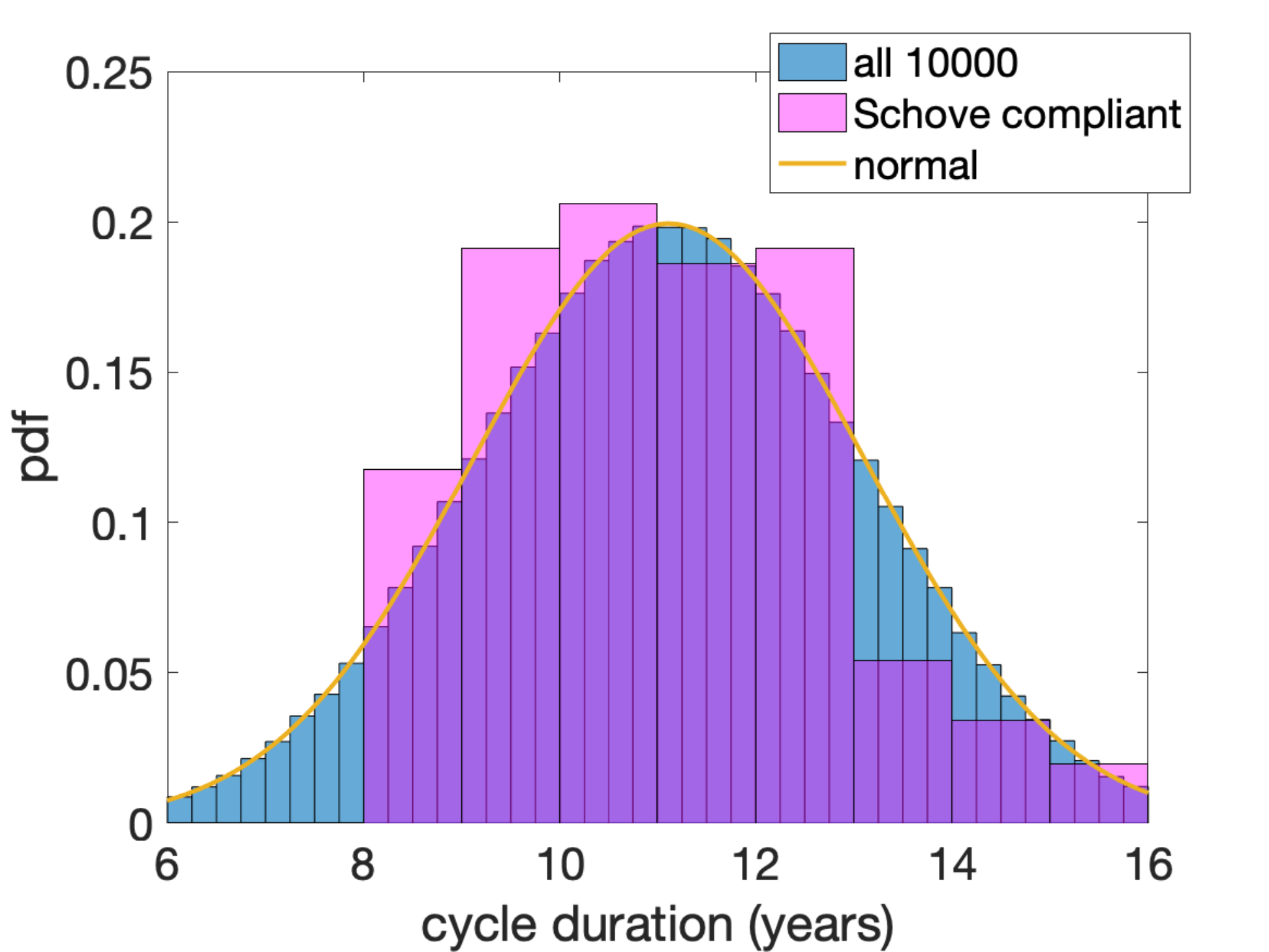}
               \hspace*{-0.03\textwidth}
               \includegraphics[width=0.415\textwidth,clip=]{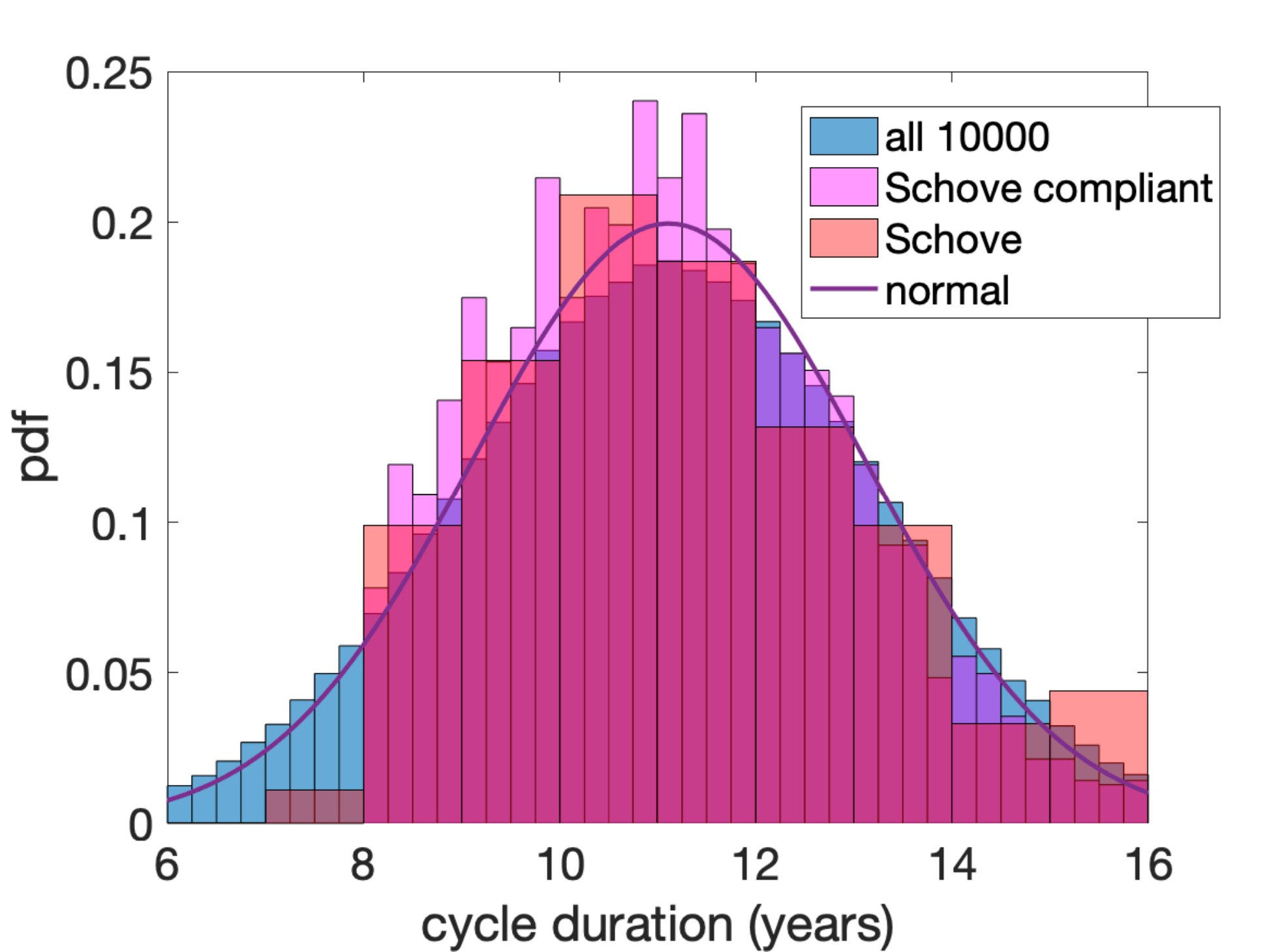}
              }
     \vspace{-0.29\textwidth}   
     \centerline{\bf     
      \hspace{0.16 \textwidth}  \color{black}{(a)}
      \hspace{0.34\textwidth}  \color{black}{(b)}
         \hfill}
     \vspace{0.26\textwidth}    
\caption{Probability density function of the duration of synthetic solar cycles. (a) Random-walk synthetics: gaussian distribution of all 10\,000 realizations (teal), and pdf of the 3 realizations that comply with Schove's assumptions (magenta). (b) Clocked synthetics: nearly gaussian distribution of the 10\,000 realizations (teal), pdf of the 42 Schove-compliant realizations (magenta), and pdf of Schove cycle durations (red). 
        }
   \label{F-pdf}
   \end{figure}

Figure \ref{F-pdf} displays the probability density function (pdf) obtained with 10\,000 realizations, for both the random-walk series (Figure \ref{F-pdf}a) and the clocked series (Figure \ref{F-pdf}b).
The pdf of Schove series is also drawn in Figure \ref{F-pdf}b.

\subsection{The impact of Schove's assumptions}
\label{S-impact}

I have examined which of the 10\,000 realizations of both families comply with the two assumptions of \citet{schove1955} recalled in Figure \ref{F-Schove}.
I find only 3 random-walk realizations, and 42 clocked realizations.
In other words, the {\bf assumptions} used by \citet{schove1955} practically exclude random variations of the duration between solar maxima.

The cycle duration pdf of Schove-compliant realizations are shown in Figure \ref{F-pdf}, while their time series and deviations from a linear fit are displayed in Appendix \ref{A-deviations} together with those of Schove's series.

  \begin{figure}    
   \centerline{\hspace*{0.015\textwidth}
               \includegraphics[width=0.7\textwidth,clip=]{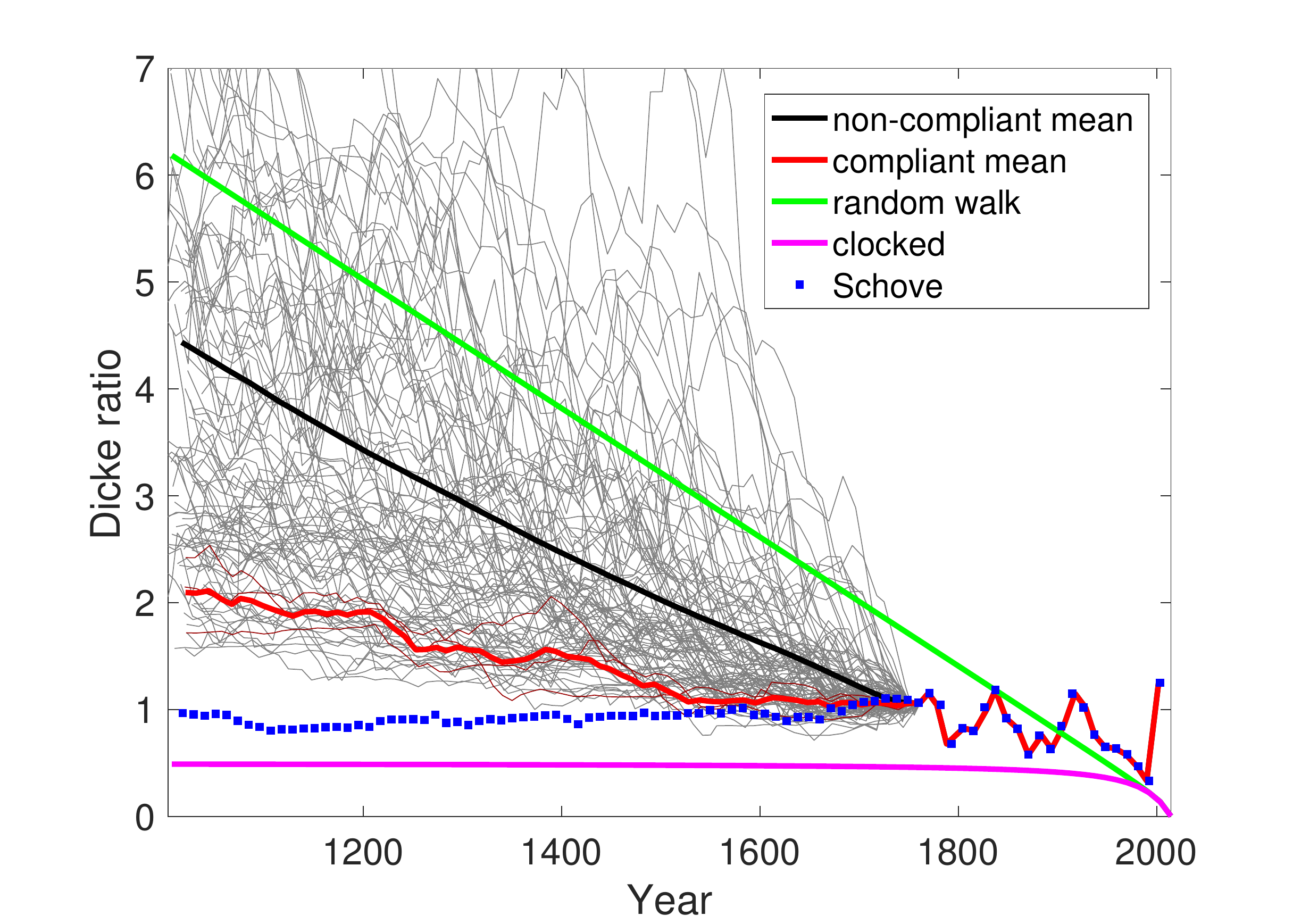}
               }
     \vspace{-0.48\textwidth}   
     \centerline{\bf     
      \hspace{0.16 \textwidth}  \color{black}{(a)}
         \hfill}
     \vspace{0.435\textwidth}    

   \centerline{\hspace*{0.015\textwidth}
               \includegraphics[width=0.7\textwidth,clip=]{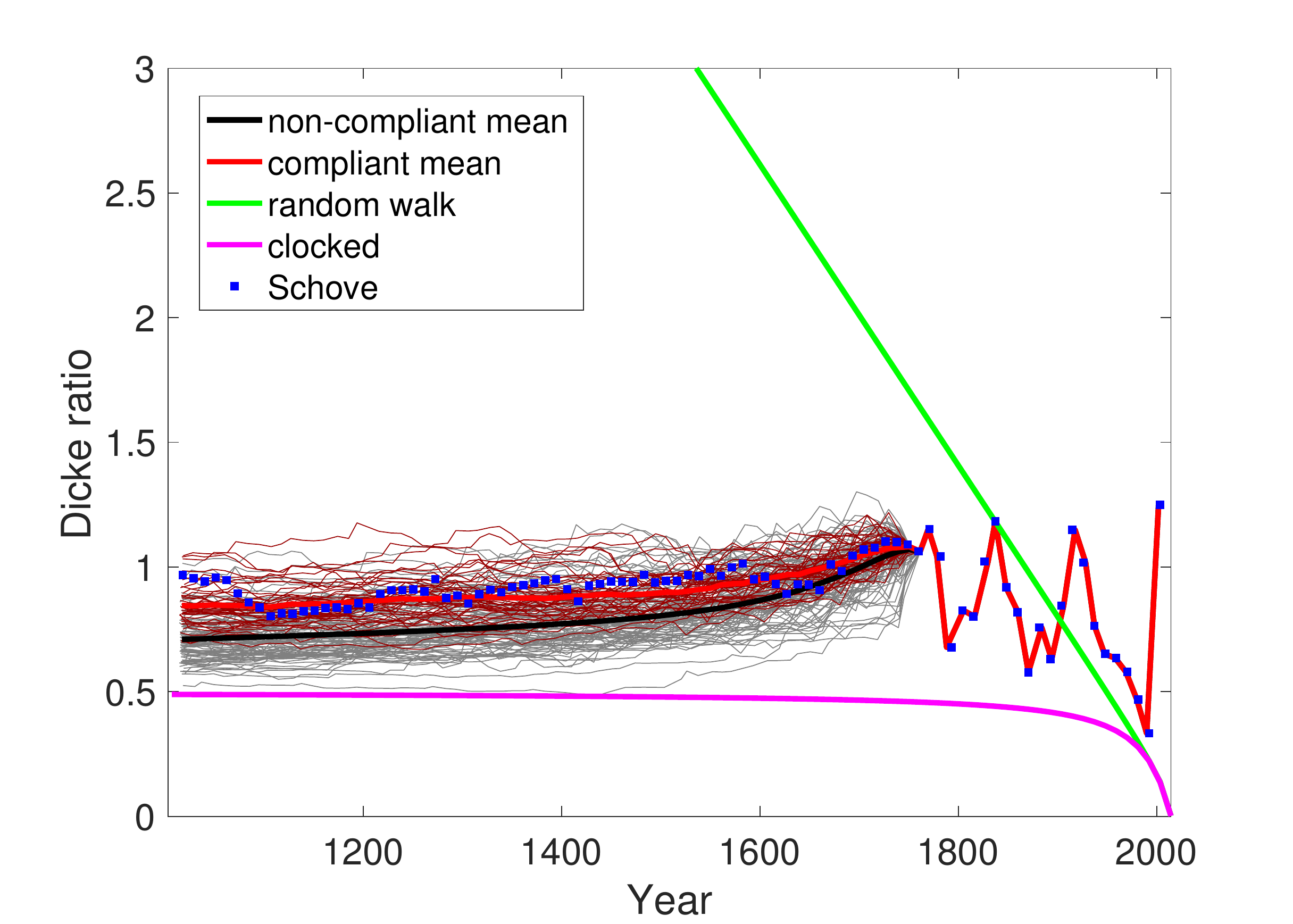}
              }
     \vspace{-0.48\textwidth}   
     \centerline{\bf     
      \hspace{0.16 \textwidth}  \color{black}{(b)}
         \hfill}
     \vspace{0.435\textwidth}    
\caption{Dicke's ratios of (a) random-walk and (b) clocked synthetic solar cycles. Random selection of 100 realizations (grey), mean of all 10\,000 realizations (thick black), Schove-compliant realizations (red) and their mean (thick red), Schove time series (blue squares). The thick green and magenta lines show the expectation of Dicke's ratio for a random-walk (equation \ref{E-random}) and for a clocked law (equation \ref{E-clock}), respectively. Note that all synthetics share the same post-1755 reliable time series.
        }
   \label{F-Dicke}
   \end{figure}

Figure \ref{F-Dicke} shows Dicke's ratios\footnote{Note that all Dicke's ratios are calculated backwards in time as in \citet{stefani2019}, since more recent dates are considered more reliable.} for a random selection of 100 realizations of both families.
The thick black line shows the mean of Dicke's ratio for all realizations, while the thick red line is the mean of the Schove-compliant realizations.
Note that Dicke's ratio of the rare Schove-compliant random-walk series is much lower that the mean of all realizations, and that individual realizations can plot far from the mean.
   
\section{Quasi-clocked magnetohydrodynamic instabilities} 
      \label{S-instabilities}

The original goal of \citet{dicke1978} was to test the compatibility of solar cycle time series with an `eruption hypothesis' expressed by \citet{kiepenheuer1959} as `each cycle represents an independent eruption of the Sun which takes about 11 years to die down'.
Fluid dynamics and magnetohydrodynamics instabilities do not necessarily behave that way, even when strong turbulence is present.
For example, quasi-periodic magnetic oscillations have been reported in the VKS dynamo experiment \citep{berhanu2009} at Reynolds numbers above $10^7$.
Nonetheless, we have seen that Dicke's ratio yields a more stringent measure of the clocked behaviour of a time series than provided by visual inspection or pdf.

\begin{figure}    
    \centerline{\includegraphics[width=0.7\textwidth,clip=]{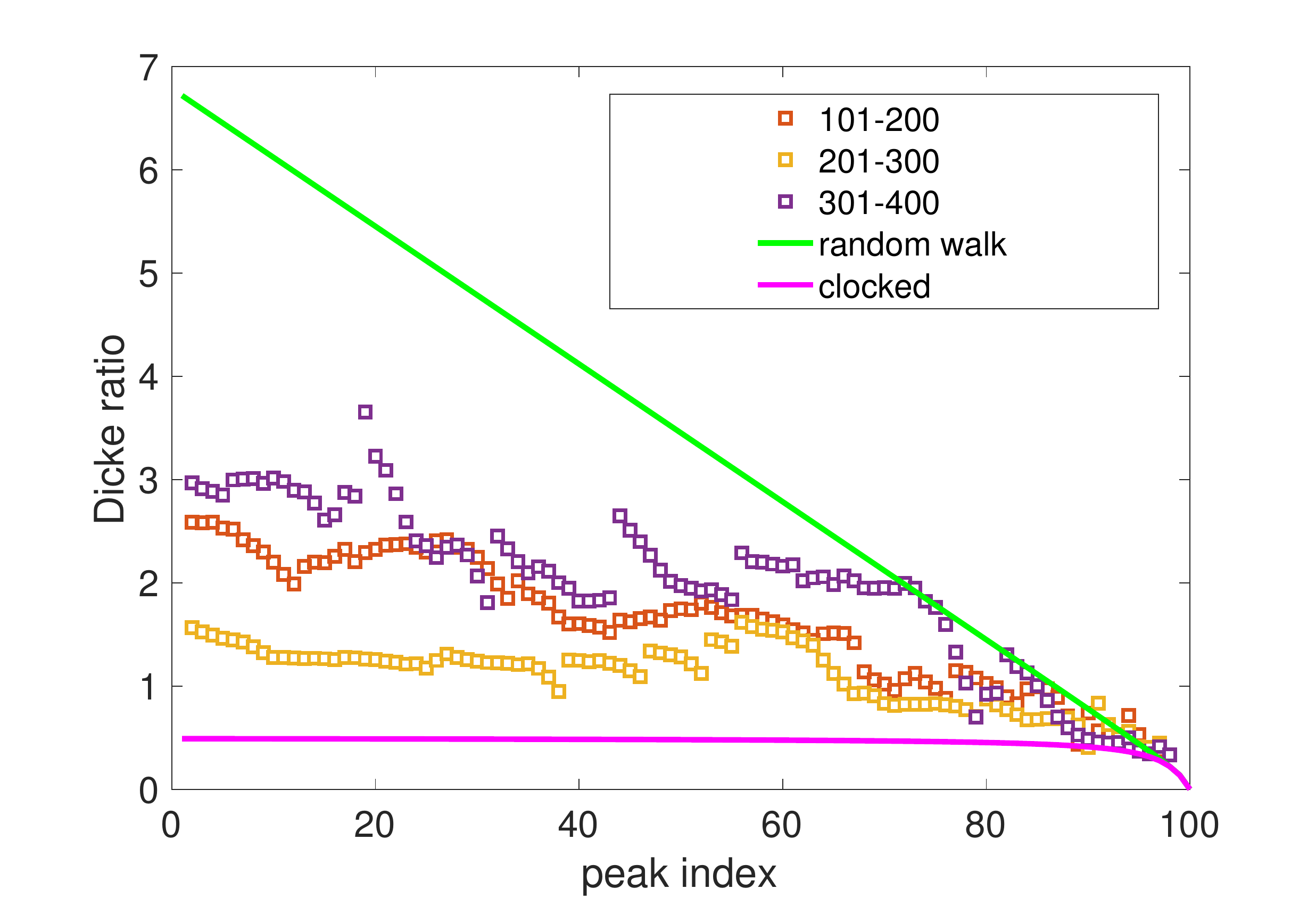}}
              \caption{Dicke's ratio of magnetic fluctuations in the DTS$\Omega$ liquid sodium experiment (see Appendix \ref{A-DTS} for details).
             Three consecutive series of 100 maxima are plotted (peaks 101 to 200, 201 to 300, and 301 to 400).
             The green and magenta lines show the expectation of Dicke's ratio for a random-walk (equation \ref{E-random}) and for a clocked law (equation \ref{E-clock}), respectively.  
                      }
   \label{F-DTS-Dicke}
   \end{figure}

The Grenoble DTS$\Omega$ liquid sodium experiment exhibits magnetic fluctuations that are often quite regular \citep{schmitt2008}.
From one such experiment, I have extracted time series of maximum induced magnetic intensity (see Appendix \ref{A-DTS} for details), and computed Dicke's ratio of several sequences of 100 consecutive maxima.
Figure \ref{F-DTS-Dicke} shows that the behaviour of these magnetic fluctuations is quasi-clocked, even though the Reynolds number is $\simeq 8 \times 10^6$ and the standard deviation of the fluctuations about $30\%$.

\section{Conclusion} 
      \label{S-conclusion} 

The demonstration by \citet{stefani2019} of a clocked behaviour for solar cycles is invalid because the 1000 years long time series they use \citep{schove1955} is {\bf clocked by construction}.
The astrological quest for a link between solar cycles and planetary tides remains as unfounded as ever.
Magnetohydrodynamics instabilities can produce quasi-periodic fluctuations that appear as almost clocked.

\begin{acks}
I thank Andr\'e Giesecke and Frank Stefani for providing clarifications on their computation of Dicke's ratio. I thank my colleagues of the geodynamo team of ISTerre for useful discussions and encouragements, and an anonymous reviewer for helpful suggestions.
This article is dedicated to Emile Okal and to the memory of Don L. Anderson.

\end{acks}



 



\begin{materialsavailability}
All matlab scripts and data used to produce the figures of this article are available as supplementary material.
\end{materialsavailability}


\begin{conflict}
The author declares that he has no conflicts of interest.
\end{conflict}
  
\appendix   

\section{Frequency spectra of synthetic tidal forcings}
\label{A-tide}

I have built synthetic tidal forcings to illustrate the lack of evidence for a ~11.2 years periodicity, as demonstrated by \citet{okal1975}.
I model the tidal forcing exerted by Jupiter, Venus, Earth and Mercury, assuming circular orbits, and starting at a time when all planets are aligned.
Mass, distances and orbital periods are taken from https://nssdc.gsfc.nasa.gov/planetary/factsheet/.

Tidal forcing of this simplified planetary system at a given meridian on the Sun can be expressed as:

\begin{equation}
Tide(t) = \sum_{p} {\frac{m_p}{d_p^3} \left[ \cos^2 \left( 2 \pi \frac{t}{T_S} -2 \pi \frac{t}{T_p} \right) -\frac{1}{3} \right]},
\label{E-tide}
\end{equation}
where $m_p$ and $d_p$ are the mass and distance from the Sun of planet $p$, respectively.
$T_p$ is its orbital period, while $T_S$ is the duration of a solar day, taken as $27$ days.

\begin{figure}    
   \centerline{\includegraphics[width=1.0\textwidth,clip=]{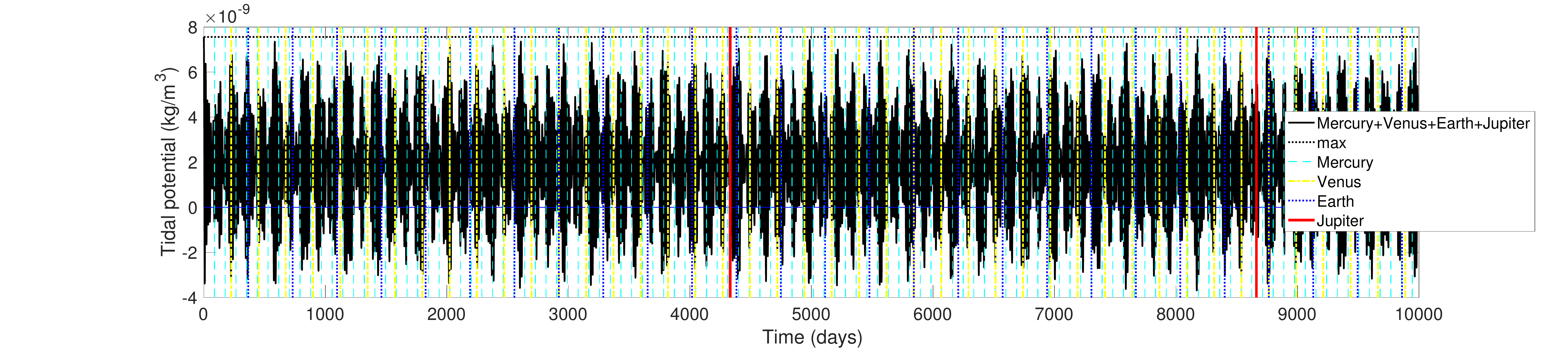}}
        \vspace{-0.22\textwidth}   
     \centerline{\bf     
      \hspace{0.04 \textwidth}  \color{black}{(a)}
         \hfill}
     \vspace{0.175\textwidth}    
    \centerline{\includegraphics[width=1.0\textwidth,clip=]{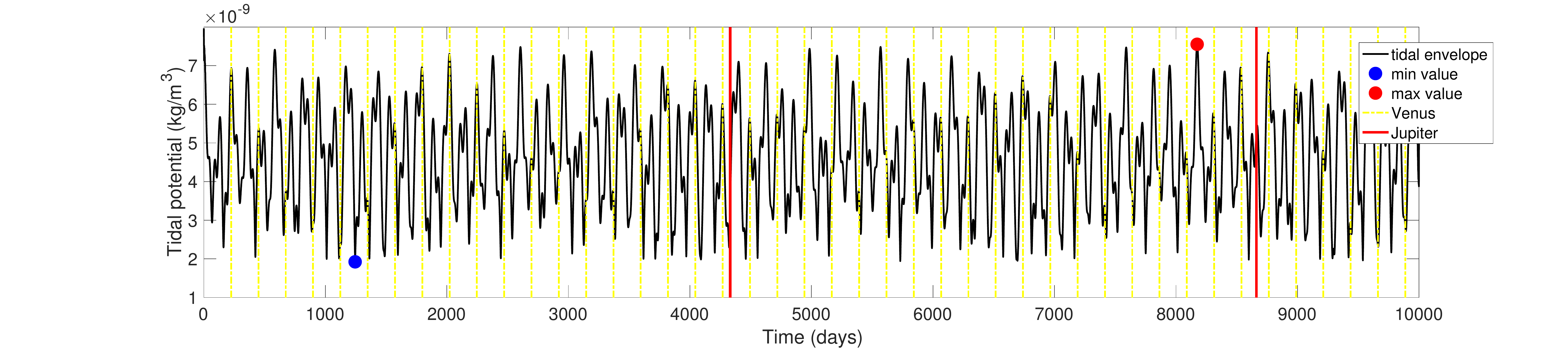}}
        \vspace{-0.22\textwidth}   
     \centerline{\bf     
      \hspace{0.04 \textwidth}  \color{black}{(b)}
         \hfill}
     \vspace{0.176\textwidth}    
    \centerline{
          \includegraphics[width=0.3\textwidth,clip=]{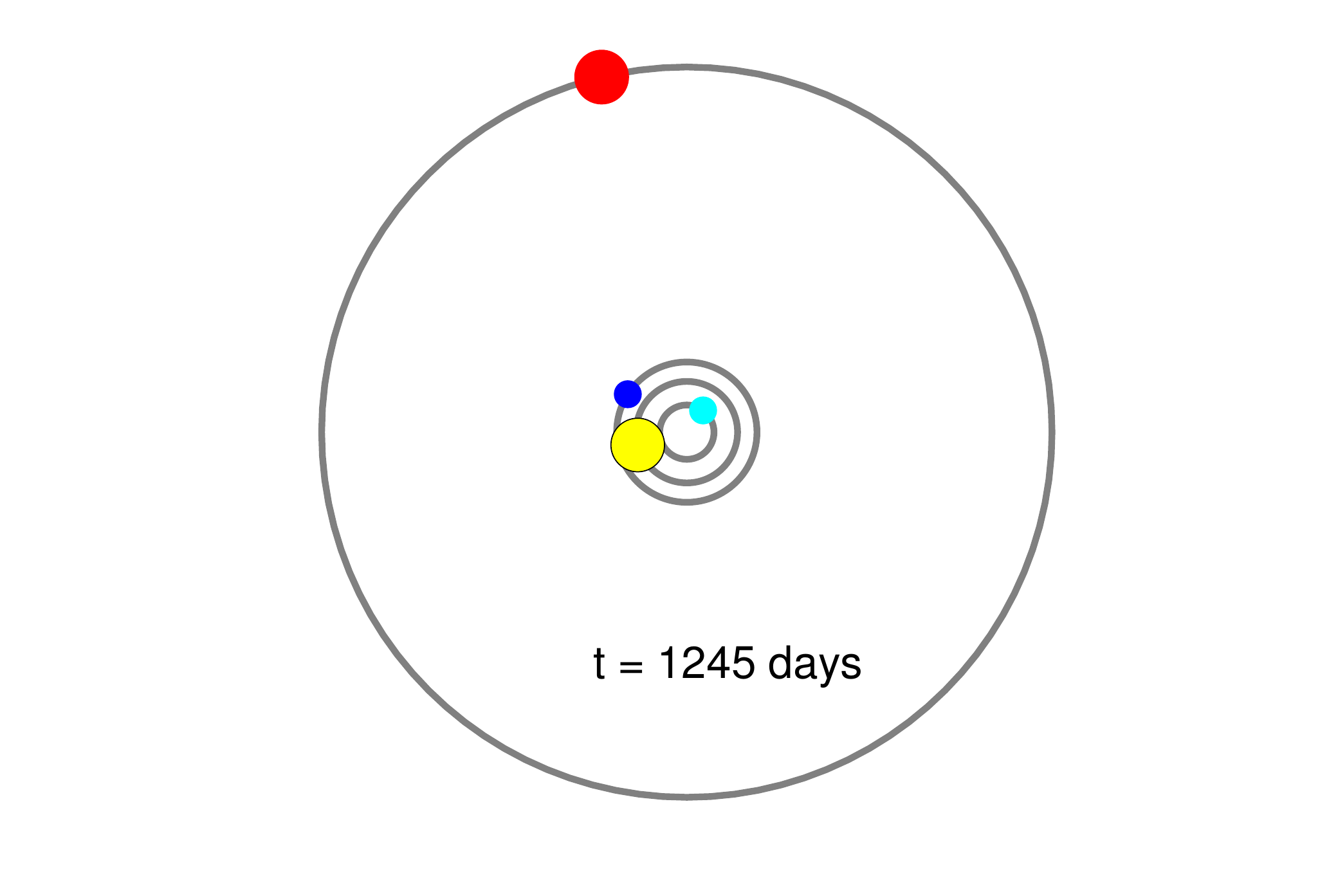}
          \hspace{1cm}
          \includegraphics[width=0.3\textwidth,clip=]{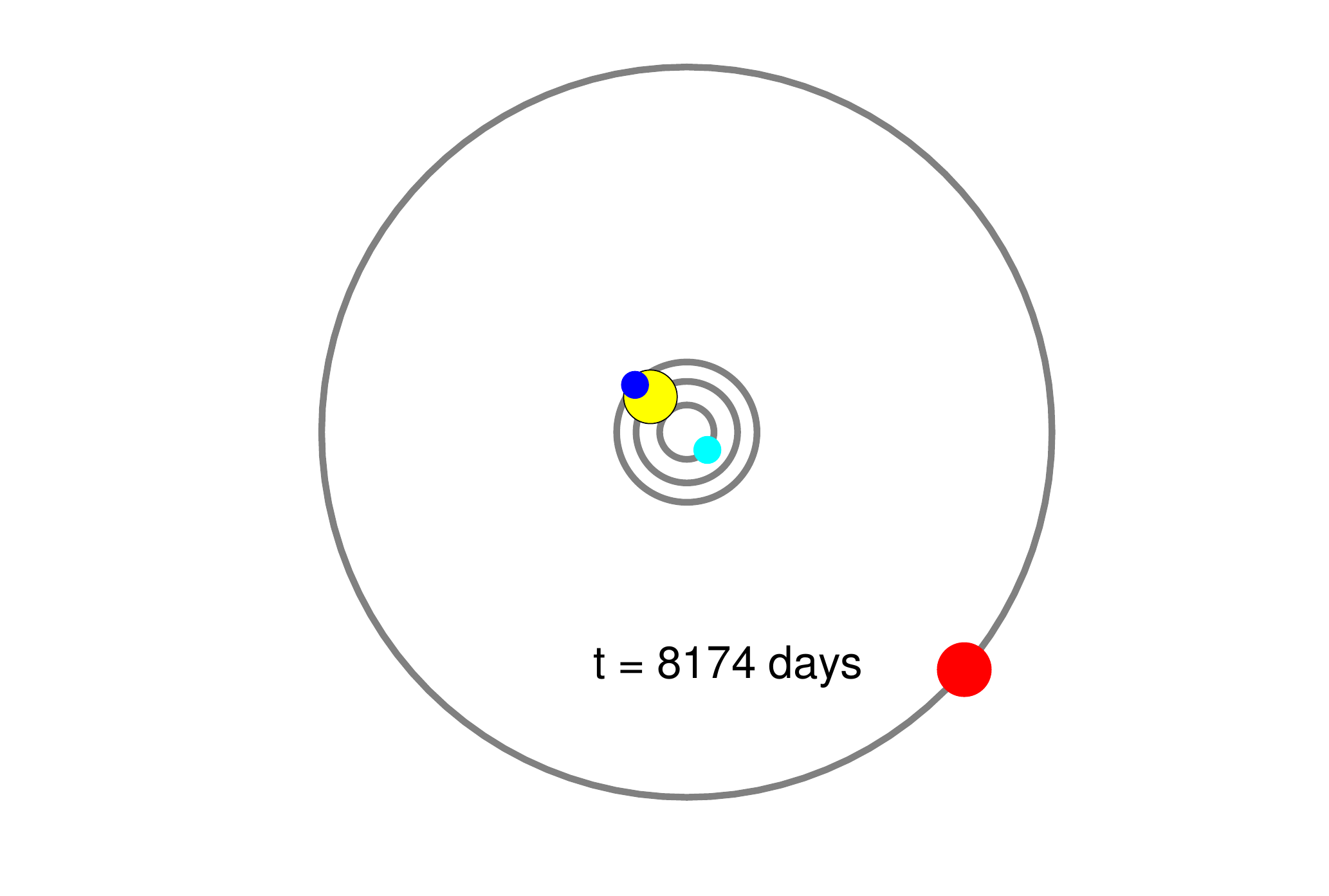}
			}
			
     \vspace{-0.21\textwidth}   
     \centerline{\bf     
      \hspace{0.16\textwidth}  \color{black}{(c)}
      \hspace{0.33\textwidth}  \color{black}{(d)}
         \hfill}
     \vspace{0.18\textwidth}    

              \caption{Time series and envelope of tidal forcing as a function of time in a realistic synthetic solar system. (a) Tidal forcing at a given meridian on the Sun, calculated from equation \ref{E-tide} over 10\,000 days (27.4 years).
              The periods of the 4 planets are marked by vertical lines of different colors.
              The maximum tide (achieved at $t=0$) is shown by a horizontal dashed black line.
              (b) envelope of the tidal forcing.
              The minimum and (second) maximum are marked by a red dot and a blue dot, respectively.
              The periods of Jupiter and Venus are shown by vertical lines.
              (c) Positions of the 4 planets at the minimum tide ($t=1245$ days).
              (d) Positions of the 4 planets at the (second) maximum tide ($t=8174$ days).
                      }
   \label{F-tide}
   \end{figure}

Figure \ref{F-tide}a shows the predicted tidal signal over a period of 10\,000 days (27.4 years).
The orbital period of Jupiter (11.86 years) is indicated by vertical red bars.
The maximum tidal forcing achieved at $t=0$ is marked by a horizontal dashed line.
It can be seen that forcings almost as large occur many times during one orbital period of Jupiter, typically each time Jupiter and Venus are aligned with the Sun.

Tidal minima are better seen in Figure \ref{F-tide}b, which displays the envelope of the tidal signal.
The minimum of the series (at $t=1245$ days) is marked by a blue dot, while the second maximum (at $t=8174$ days) is shown by a red dot.
The corresponding positions of the 4 planets are given in Figures \ref{F-tide}c and  \ref{F-tide}d; respectively. As expected, all 4 planets are almost aligned with the Sun at the maximum, while Jupiter cancels Venus' tide and Mercury cancels the Earth's tide at the minimum.

\begin{figure}    
   \centerline{\includegraphics[width=0.5\textwidth,clip=]{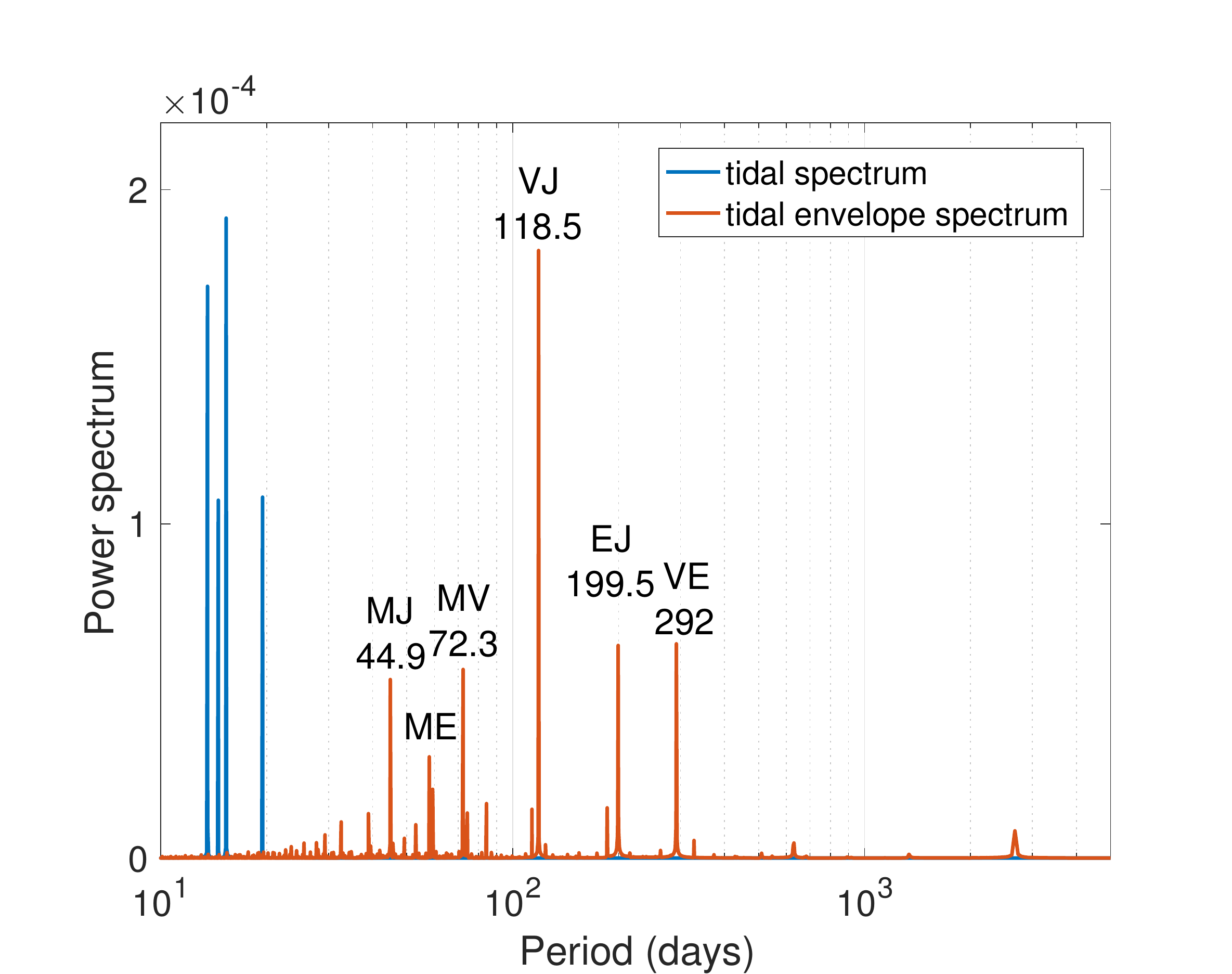}
              }
              \caption{Power spectra of the tidal signal (blue) and of its envelope (orange), as a function of period.
              Tidal peaks of the envelope spectrum appear at syzygies of pairs of planets with the Sun.
              Following \citet{okal1975}, I label them with the initials of the two planets.
              The largest tide is `VJ' at 118.5 days when Venus and Jupiter are aligned with the Sun.
                      }
   \label{F-spectra}
   \end{figure}

Figure \ref{F-spectra} presents the spectra of a 360 years-long synthetic tidal signal (blue) and of its envelope (orange).
The 4 peaks of the former simply correspond to half a solar day as seen from each planet (see equation \ref{E-tide}).
The spectrum of the envelope is dominated by peaks at the periods of syzygies of pairs of planets with the Sun, and their overtones.
The spectrum is almost flat for periods beyond 300 days.
This plot mimics Figure 3 of \citet{okal1975}, which shows the full tidal spectrum, `taking into account the complete orbital elements [of the 4 planets], including eccentricity, inclination and their variation with time' over a period of 1\,800 years.
Orbital eccentricity adds up small tidal peaks at the orbital periods of Jupiter (11.86 years) and Mercury (0.241 years), but nothing shows up at 11.2 years (4088 days), as emphasized by \citet{okal1975}.

\section{Time series and deviations of synthetic solar maxima}
\label{A-deviations}

\begin{figure}    
   \centerline{\hspace*{0.015\textwidth}
               \includegraphics[width=0.515\textwidth,clip=]{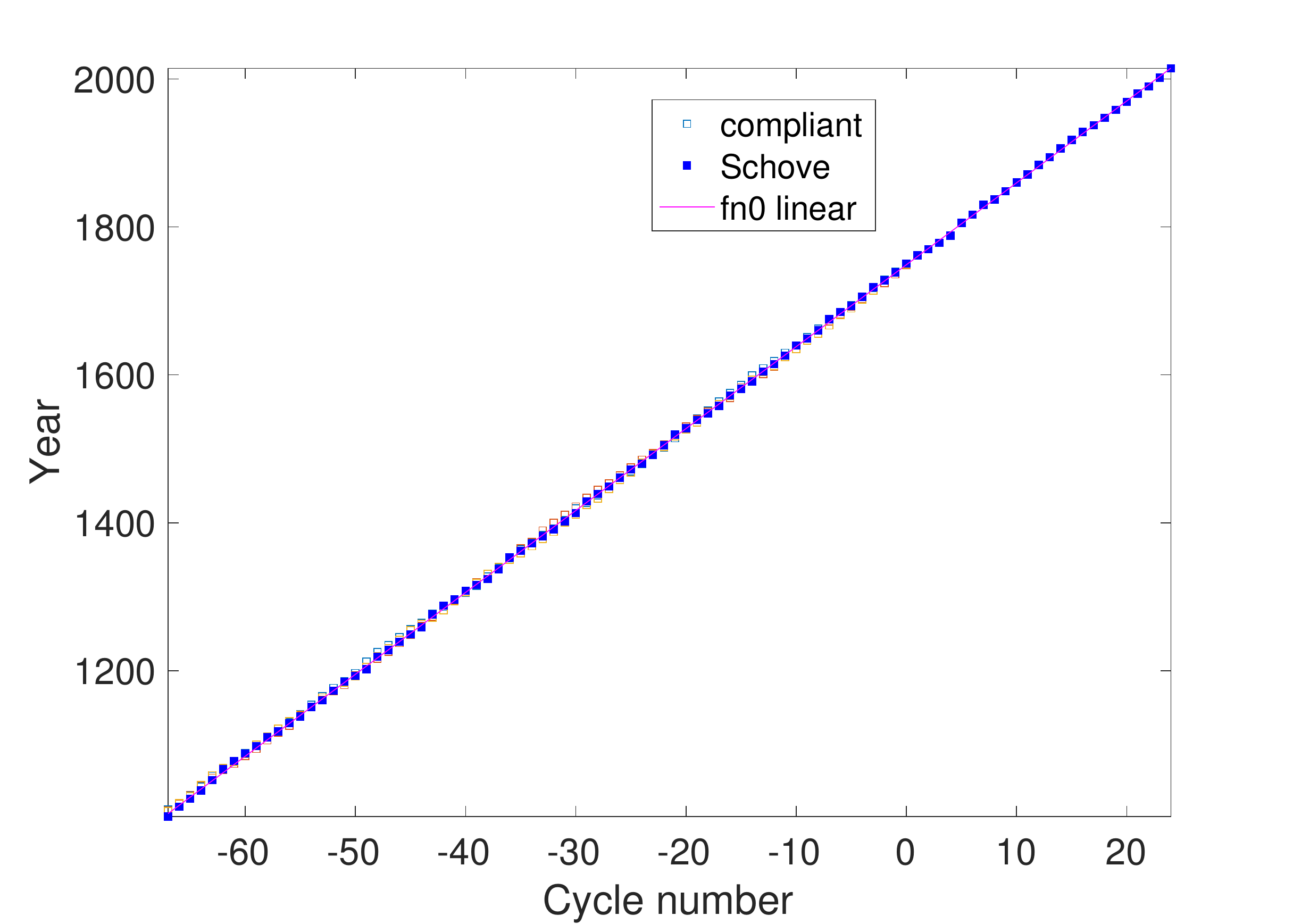}
               \hspace*{-0.03\textwidth}
               \includegraphics[width=0.515\textwidth,clip=]{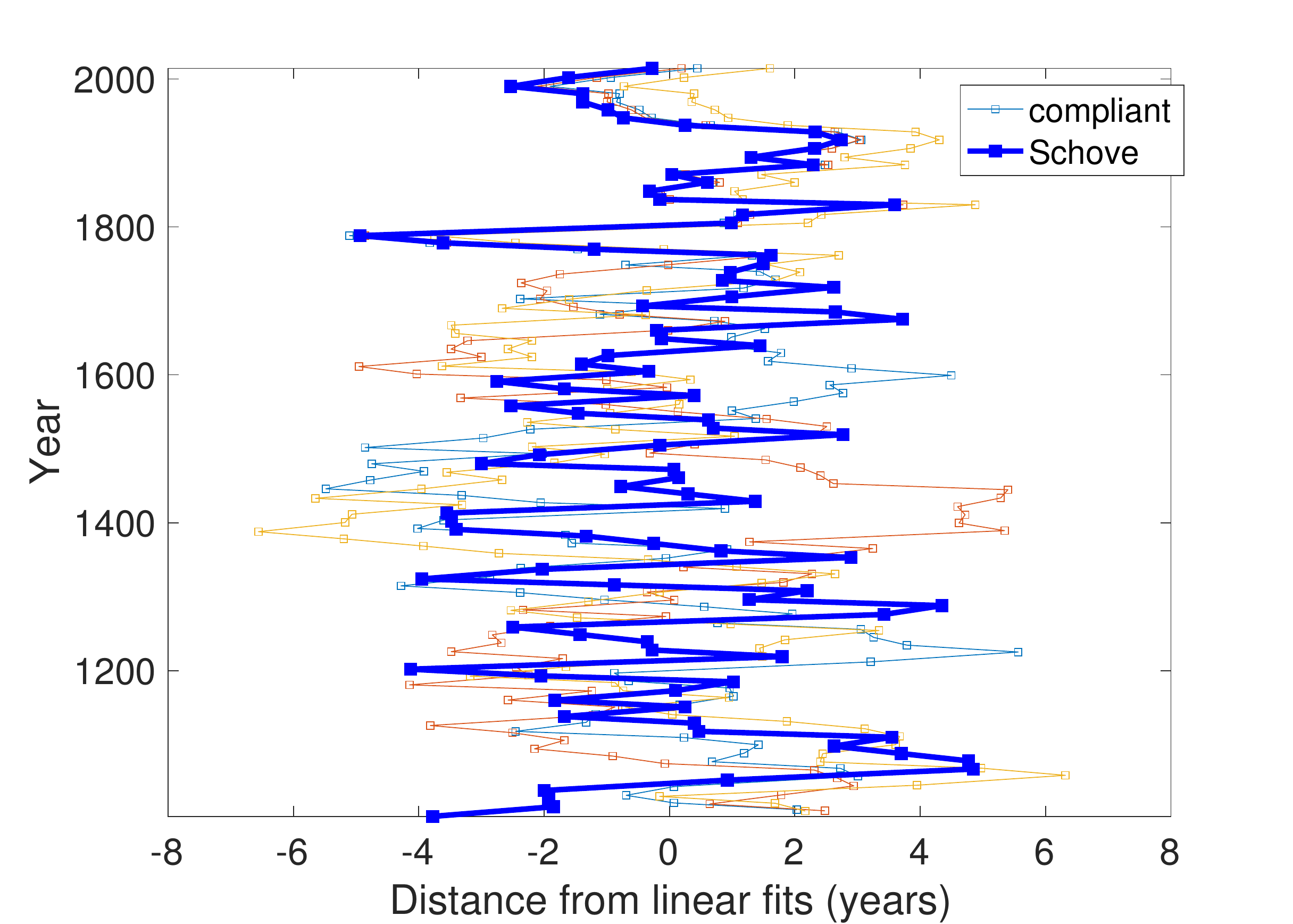}
              }
     \vspace{-0.34\textwidth}   
     \centerline{\bf     
      \hspace{0.07 \textwidth}  \color{black}{(a)}
      \hspace{0.44\textwidth}  \color{black}{(b)}
         \hfill}
     \vspace{0.3\textwidth}    
\caption{(a) Time series and (b) deviations of Schove-compliant random-walk synthetic solar cycles.
The blue squares are for Schove's maxima before 1755, and from \citet{sidc2022} after 1755.
        }
   \label{F-deviations-random}
   \end{figure}

\begin{figure}    
   \centerline{\hspace*{0.015\textwidth}
               \includegraphics[width=0.515\textwidth,clip=]{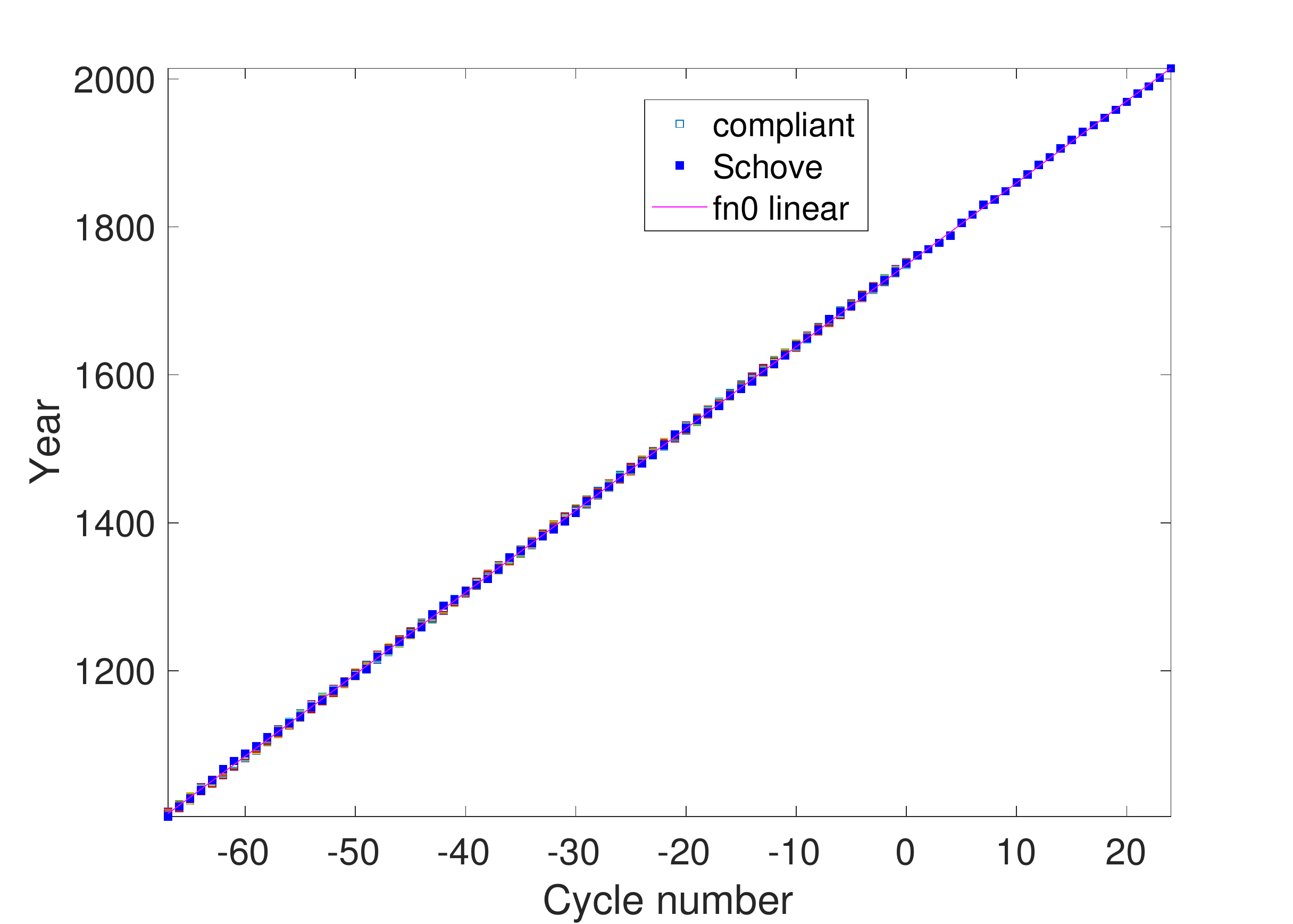}
               \hspace*{-0.03\textwidth}
               \includegraphics[width=0.515\textwidth,clip=]{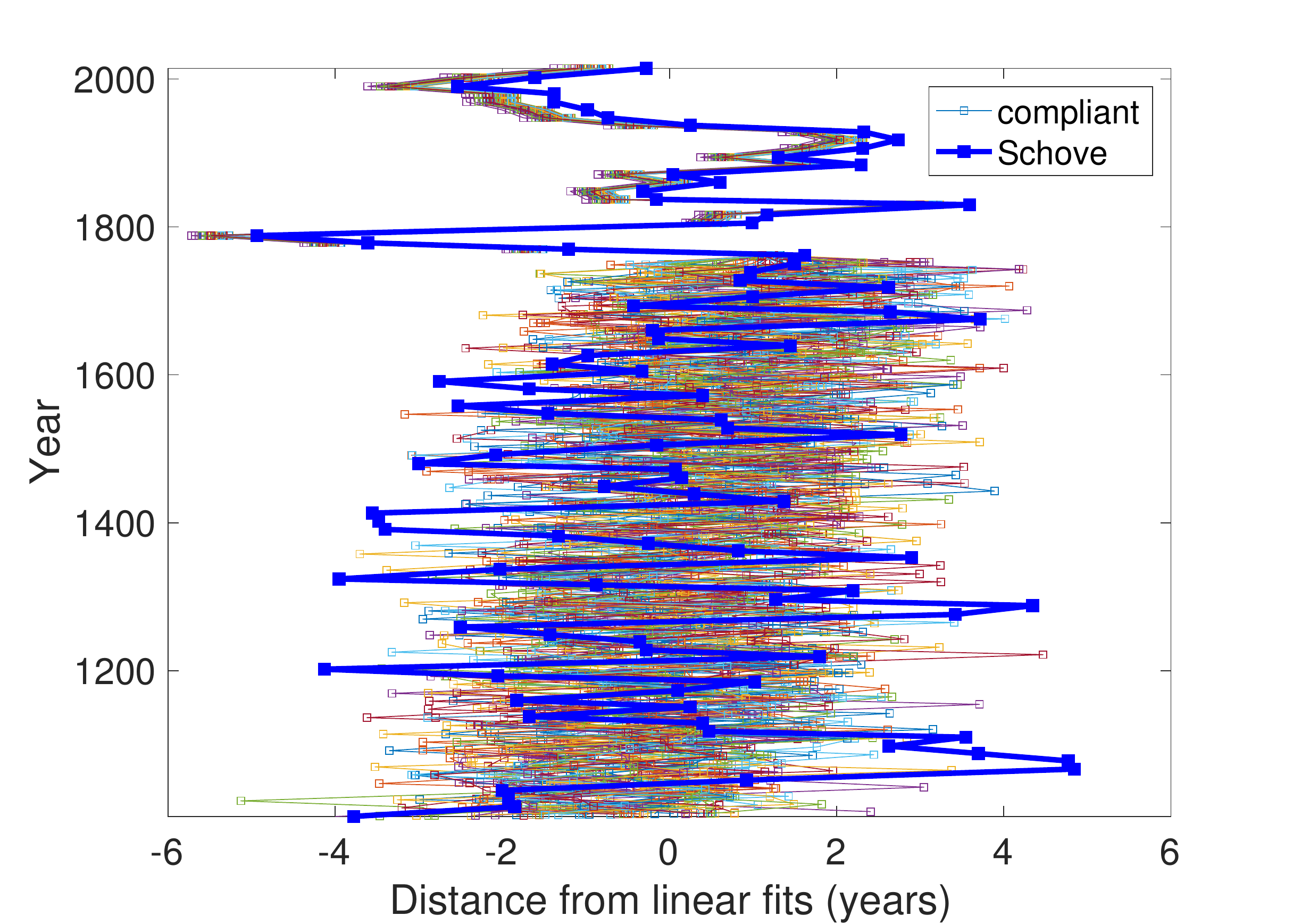}
              }
     \vspace{-0.34\textwidth}   
     \centerline{\bf     
      \hspace{0.07 \textwidth}  \color{black}{(a)}
      \hspace{0.44\textwidth}  \color{black}{(b)}
         \hfill}
     \vspace{0.3\textwidth}    
\caption{(a) Time series and (b) deviations of Schove-compliant clocked synthetic solar cycles. The blue squares are for Schove's maxima before 1755, and from \citet{sidc2022} after 1755.
        }
   \label{F-deviations-clock}
   \end{figure}

Figures \ref{F-deviations-random} and \ref{F-deviations-clock} display the time series and deviations of the synthetic solar cycles that comply with Schove's assumptions, to compare with Figure 1 of \citet{stefani2019}. 
Deviations of each realization are the difference between dates of maxima and a linear fit of these dates.

\section{Dicke's ratio of magnetic fluctuations in the DTS$\Omega$ experiment}
\label{A-DTS}

The DTS$\Omega$ experiment was built to study magnetohydrodynamics in the magnetostrophic regime, in which Lorentz and Coriolis forces are dominant.
Fifty liters of liquid sodium are enclosed in a spherical container that can rotate around a vertical axis.
An inner central sphere can rotate independently around the same axis, and hosts a strong permanent magnet producing an axial dipolar magnetic field.
The 3 components of the induced magnetic field are measured at the surface of the outer sphere at 20 equally-spaced latitudes from  $-57^\circ$ to  $57^\circ$ (see \citet{schmitt2013} for more details).
The frequency spectra of electric and magnetic fluctuations reveal a quasi-periodic behaviour that can be linked to the presence of magneto-Coriolis waves \citep{schmitt2008,schmitt2013} or instabilities \citep{figueroa2013,kaplan2018}.

\begin{figure}    
   \centerline{
   \hspace{0.01 \textwidth}
   \includegraphics[width=1.02\textwidth,clip=]{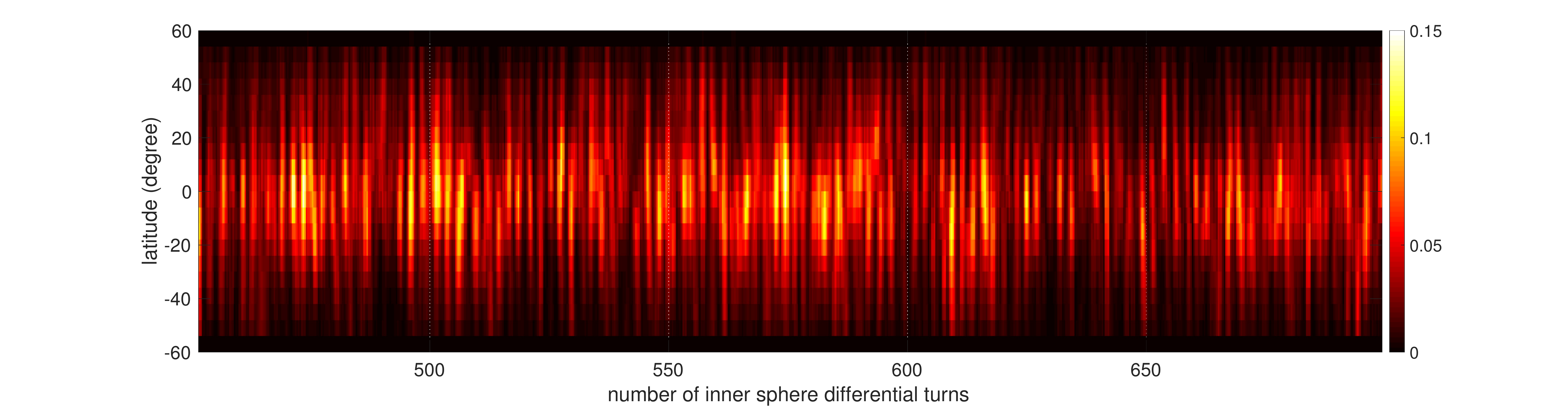}
   }
     \vspace{-0.26\textwidth}   
     \centerline{\bf     
      \hspace{0.04 \textwidth}  \color{black}{(a)}
         \hfill}
     \vspace{0.215\textwidth}    

    \centerline{\includegraphics[width=1.0\textwidth,clip=]{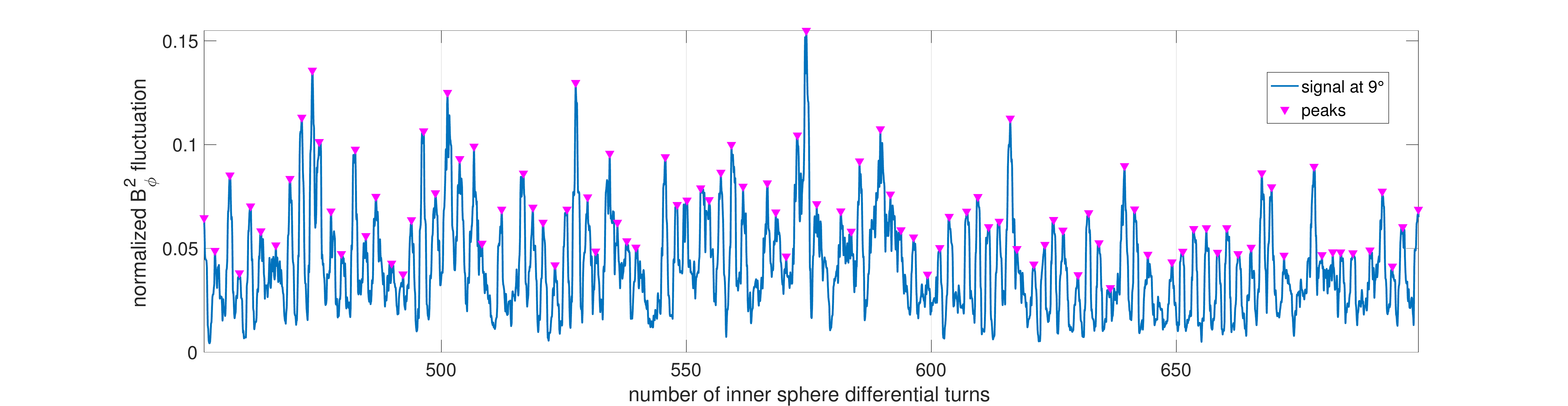}}

     \vspace{-0.26\textwidth}   
     \centerline{\bf     
      \hspace{0.04 \textwidth}  \color{black}{(b)}
         \hfill}
     \vspace{0.22\textwidth}    

              \caption{Magnetic fluctuations in the DTS$\Omega$ experiment.
              (a) Color-coded image of the square of azimuthal magnetic field fluctuations, normalized by the square of the imposed magnetic field (\%) at the surface of the outer sphere.
              The magnetic field is measured at the 20 latitudes which form the y-axis.
              The x-axis is time given in number of turns of the inner sphere with respect to the outer sphere.
              The records are averaged over one such differential turn.
              (b) Extraction of the same signals at a latitude of $9^\circ$, showing a succession of peaks labelled by magenta triangles.
                      }
   \label{F-DTS}
   \end{figure}

An example of such quasi-periodic magnetic fluctuations is given in Figure \ref{F-DTS}.
The spin rate of the outer sphere is $f_o \simeq 10$ Hz, while the differential rotation of the inner sphere is $\Delta f \simeq -20$ Hz, yielding fluid velocities $U \simeq 25$ m/s.
With an outer radius $r_o = 0.21$ m, the Reynolds number $Re = \frac{U r_o}{\nu} \simeq 8 \times 10^6$ and the magnetic Reynolds number $Rm =  \frac{U r_o}{\eta} \simeq 60$, with $\nu$ the kinematic viscosity, and $\eta$ the magnetic diffusivity.
Figure \ref{F-DTS}a displays a latitude-versus-time color-coded image of the squared azimuthal magnetic fluctuations averaged over one turn of the inner sphere, in a time-window of some 250 rotations of the inner sphere.
Latitudinally-coherent quasi-periodic fluctuations of variable intensity dominate.
Figure \ref{F-DTS}b is the time-record of the same fluctuations at a latitude of $9^\circ$.
I have extracted the 100 maxima of this record, plotted as triangles in figure \ref{F-DTS}b.
The duration between maxima is $0.11 \pm 0.03$ s.
Dicke's ratio of the resulting time series is plotted in Figure \ref{F-DTS-Dicke}, together with Dicke's ratio of the previous and next series of 100 maxima.
All three series appear to be closer to a `clocked' behaviour than to a random-walk behaviour.

\bibliographystyle{spr-mp-sola}
\bibliography{biblio_Nataf}  

\IfFileExists{\jobname.bbl}{} {\typeout{}
\typeout{****************************************************}
\typeout{****************************************************}
\typeout{** Please run "bibtex \jobname" to obtain} \typeout{**
the bibliography and then re-run LaTeX} \typeout{** twice to fix
the references !}
\typeout{****************************************************}
\typeout{****************************************************}
\typeout{}}

\end{article} 

\end{document}